\documentclass{appolb}
\usepackage{graphicx}
\usepackage{epsfig}
\begin{document}

\title{PARTICLE-HOLE NATURE OF THE LIGHT HIGH-SPIN TOROIDAL ISOMERS\thanks{
Presented at the Zakopane Conference on Nuclear Physics
"Extremes of the Nuclear Landscape", August 30 - September 7, 2014, Zakopane, Poland.} }

\author{ A.~Staszczak
\address{Institute of Physics, Maria Curie-Sk{\l}odowska University, PL-20031 Lublin, Poland} \\[0.5cm]
Cheuk-Yin~Wong
\address{Physics Division, Oak Ridge National Laboratory, Oak Ridge, TN 37831, USA} }
\maketitle

\begin{abstract}
Nuclei under non-collective rotation with a large angular 
momentum above some threshold can assume a toroidal shape. 
In our previous work, we showed by using cranked Skyrme-Hartree-Fock
approach that even-even, N=Z, high-K, toroidal isomeric 
states may have general occurrences for light nuclei with
28$\leq$$A$$\leq$52. We present here some additional
results and systematics on the particle-hole nature of 
these high-spin toroidal isomers.
\end{abstract}


\section{Introduction}
A closed orientable surface has a topological invariant known as the
Euler characteristic $\chi$=$2-2g$, where the genus $g$ is the number
of holes in the surface. Nuclei as we now know them have the topology
of a sphere with $\chi$=2. Wheeler suggested that under appropriate
conditions the nuclear fluid may assume a toroidal shape ($\chi$=0).
Using the liquid-drop model \cite{Won73} and the rigid-body moment of
inertia \cite{Won78}, it was shown that a toroidal nucleus, endowed
with an angular momentum $I$=$I_z$ aligned about its symmetry $z$-axis
beyond a threshold, is stable against the breathing deformation in
which the major radius $R$ contracts and expands. The rotating
liquid-drop nuclei can also be stable against sausage instabilities
(know also as Plateau-Rayleigh instabilities, in which the torus
breaks into smaller fragments), when the same mass flow is maintained
across the meridian to lead to high-$I_z$ isomers within an angular
momentum window \cite{Won78}.

\section{Recent Investigations on Rotating Toroidal Nuclei}

Recently Ichikawa \textit{et al.}~\cite{Ich12} found that toroidal
high-spin isomer with $I$=60$\hbar$ may be in a local energy minimum
in the excited states of $^{40}$Ca.  They used a cranked
Skyrme-Hartree-Fock (HF) method starting from the initial ring
configuration of 10 alpha particles. In Ref.~\cite{Ich14a} it was
found from the time-dependent HF and the random-phase approximation
that a collective rotation about the axis perpendicular to symmetry
axis of the toroidal isomer $^{40}$Ca($I$=60$\hbar$) can result in a
pure collective precession motion. In our previous
study~\cite{Sta14}, we found that rotating toroidal nuclei have
general occurrences and we located 18 even-even, $N$=$Z$, high-spin
toroidal isomeric states: $^{28}$Si($I$=44$\hbar$), $^{32}$S($I$=48,
66$\hbar$), $^{36}$Ar($I$=56, 72, 92$\hbar$), $^{40}$Ca($I$=60,
82$\hbar$), $^{44}$Ti($I$=68, 88, 112$\hbar$), $^{48}$Cr($I$=72, 98,
120$\hbar$), and $^{52}$Fe($I$=52, 80, 104, 132$\hbar$) in the region
16$\leq$A$\leq$52.
Subsequent to the work of \cite{Sta14}, Ichikawa
\textit{et al.} in Ref.~\cite{Ich14b} investigated the existence of
toroidal isomers and their precession motions for nuclei with
28$\leq$A$\leq$52.  They also obtained high-spin toroidal isomers in
$^{36}$Ar, $^{40}$Ca, $^{44}$Ti, $^{48}$Cr, and $^{52}$Fe,
confirming the general occurrence of high-spin toroidal isomers in
this mass region in~\cite{Sta14}.

\begin{figure}[htb]
\begin{center}
  \includegraphics[width=0.45\columnwidth]{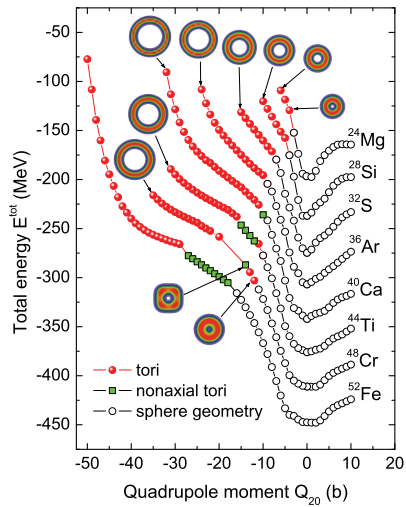}
  \caption{\label{Fig1} (Color online.) The total HFB energy of
    $^{24}$Mg, $^{28}$Si, $^{32}$S, $^{36}$Ar, $^{40}$Ca, $^{44}$Ti,
    $^{48}$Cr, and $^{52}$Fe as a function of the quadrupole moment
    $Q_{20}$ for the case of $I$=0, taken from Ref.~\cite{Sta14}.
    Axially-symmetric toroidal configurations are indicated by full
    circles, axially-asymmetric toroidal configurations by full
    squares, and configurations with a spherical topology by open
    circles. Some toroidal density distributions are displayed.}
\end{center}
\end{figure}

To study the occurrence of high-spin toroidal isomers we use a
three-step method~\cite{Sta14}, (see also~\cite{Sta08}). First, using
the Skyrme-Hartree-Fock-Bogoliubov (HFB) model with the quadrupole
moment constraints on $Q_{20}$, we look for those oblate
configurations with toroidal nuclear density distributions, as shown
in Fig.~\ref{Fig1}. The energies of axially-symmetric toroidal
configurations as a function of $Q_{20}$ lie on a slope. This
indicates that the magnitudes of the shell corrections are not
sufficient to stabilize the tori against the tendency to return to the
topology of a sphere.

\begin{figure}[htb]
\begin{center}
  \includegraphics[width=0.8\columnwidth]{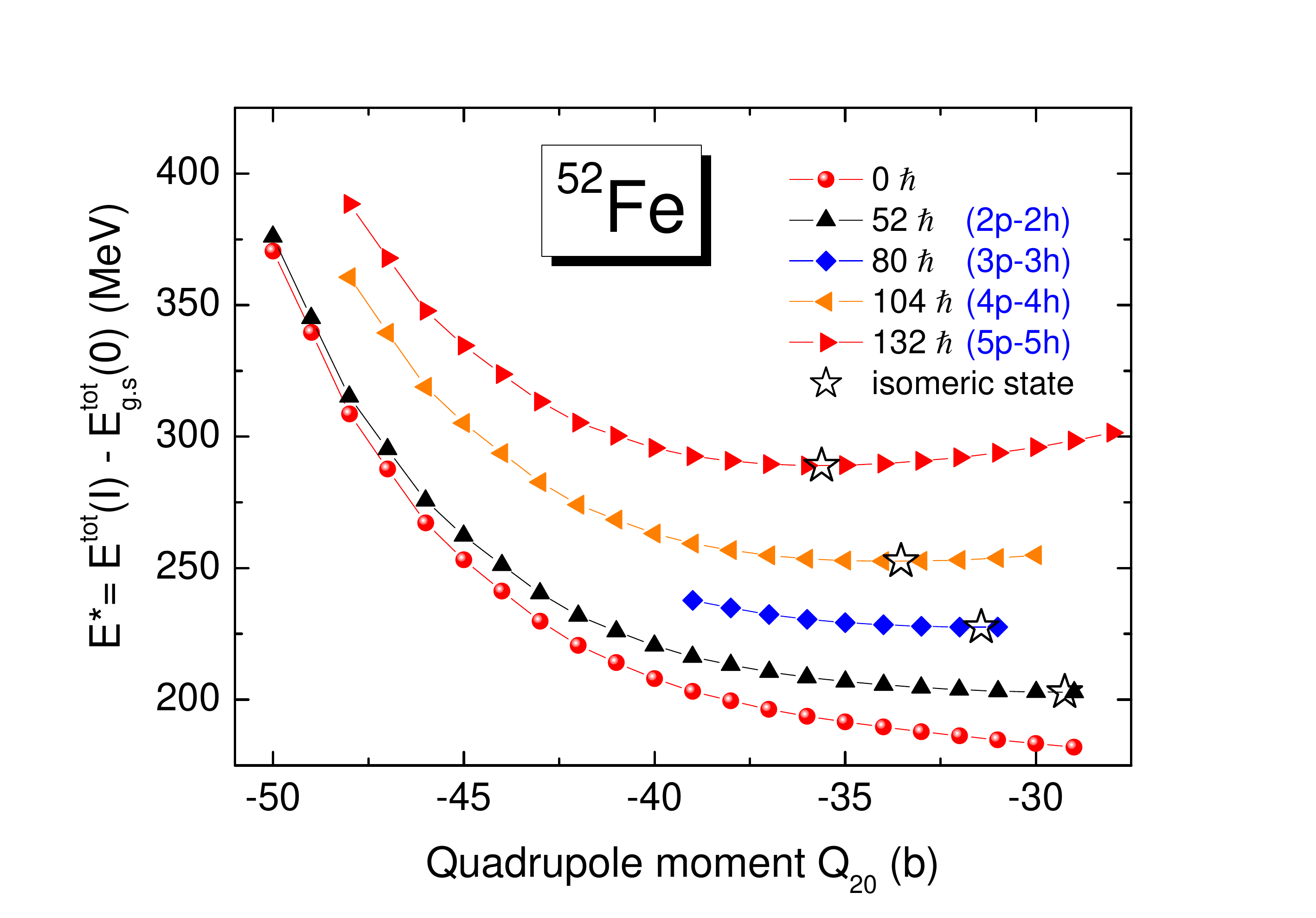}
  \caption{\label{Fig2} (Color online.) The excitation energy $E^*$ of
    high-spin toroidal states of $^{52}$Fe as a function of $Q_{20}$
    for different $I$=$I_z$ about the symmetry axis. The locations of
    isomeric toroidal energy minima are indicated by the star
    symbols.}
\end{center}
\end{figure}

We next take these toroidal configurations as the initial
configurations in $Q_{20}$-constrained cranking Skyrme-HF
calculations.  For a non-collectively rotating toroidal nucleus around
the symmetry $z$-axis with aligned angular momentum, $I$=$I_z$, we use
a Lagrange multiplier $\omega$ to describe the constraint
$I_{z}$=$\langle \hat{J}_{z} \rangle$=$\sum_{i=1}^{N} \Omega_{zi}$,
where $\Omega_{zi}$ is the $z$-component of $i$-th single-particle
total angular momentum.  When we locate the configurations which lie
close to a local minimum for each quantized value of angular momentum,
$I$=$I_z$, we repeat the cranked HF calculations without the $Q_{20}$
constraint to find the high-spin toroidal isomeric states in free
convergence in the last step. Results of this method in the case of
$^{52}$Fe are shown in Fig.~\ref{Fig2}.

\begin{figure}[htb]
\begin{center}
  \includegraphics[width=0.9\columnwidth]{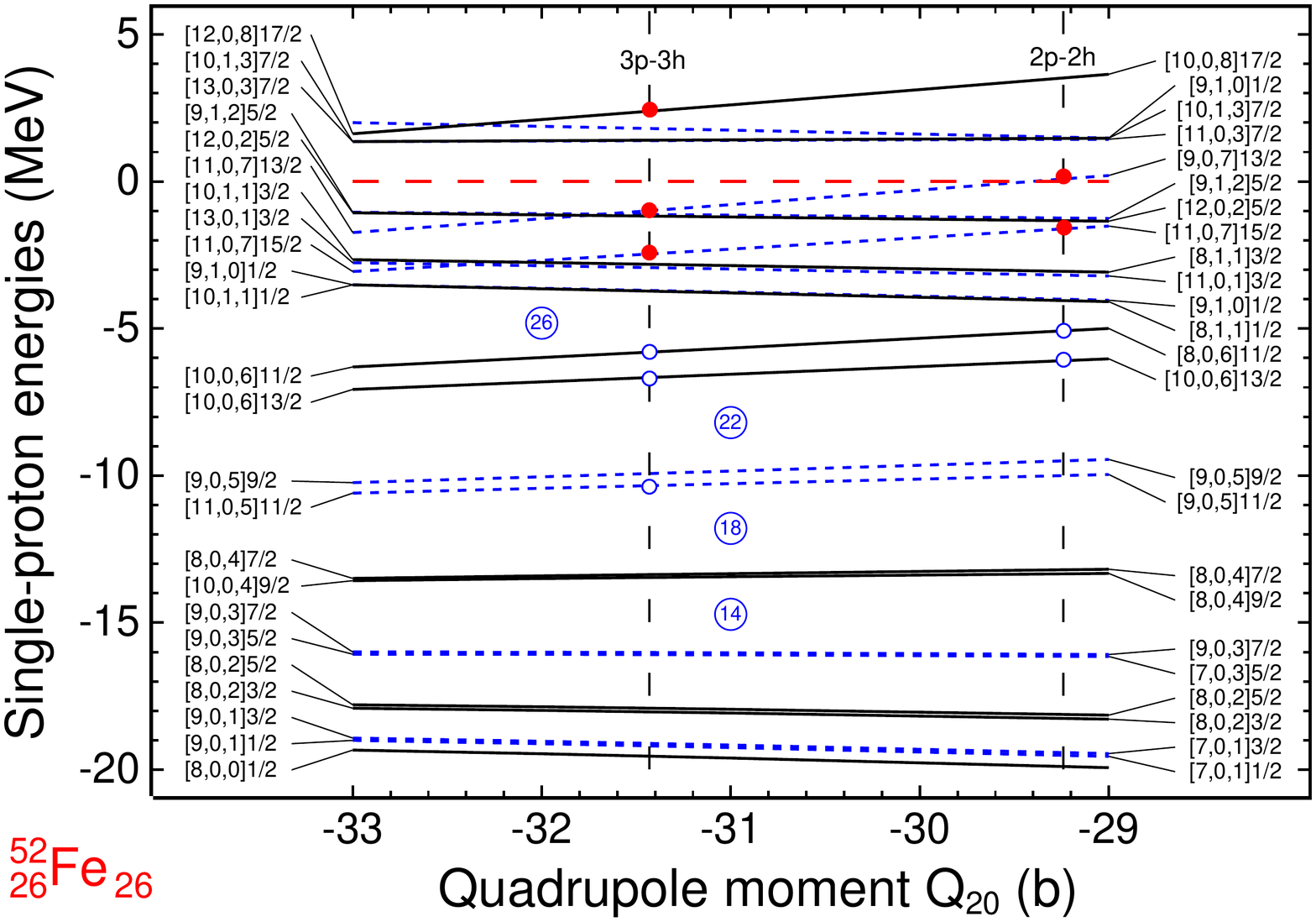}
  \caption{\label{Fig3} (Color online.) The toroidal
    proton-quasiparticle energies as a function of the quadrupole
    moment $Q_{20}$ obtained in the SkM*-HFB model for $^{52}$Fe with
    $I$=0. They are labelled by $[Nn_{z}\Lambda]\Omega$, with even
    parity levels as solid lines, and odd parity levels as dashed
    lines. Starting from the $I$=0 configuration, the 2p-2h and 3p-3h
    excitations shown in the plot (with holes as open circles and
    particles in solid circular points) lead to non-collective rotations with a
    total $I_z$=26 and 40$\hbar$, respectively. The vertical dashed lines coincide
    with the quadrupole deformations of two first toroidal isomers (Fig.~\ref{Fig2}).
    The horizontal (red) dashed line represents zero quasiparticle energy for $I$=0.}
\end{center}
\end{figure}

From the quantum mechanical point of view, the non-collective rotation 
around the symmetry axis corresponds only to particle-hole (p-h)
excitations in the axially-symmetric, $I$=0, nucleus.
We plot in Fig.~\ref{Fig3} the $I$=0 toroidal proton-quasiparticle 
state energies as a function of the quadrupole moment $Q_{20}$ obtained 
in SkM*-HFB model for $^{52}$Fe. The toroidal quasiparticle states 
are labelled by asymptotic quantum numbers $[Nn_{z}\Lambda]\Omega$. 
It is clear from Fig.~\ref{Fig3}, and also Fig.~1(a) in Ref.~\cite{Sta14}, 
that the low-lying states possess the nodal quantum number of $n_z$=0. 
The occupation numbers of 14, 18, 22, and 26 in Fig.~\ref{Fig3} indicate 
the toroidal shell-gaps, as in Fig.~1(a) of \cite{Sta14}.
The occupation of all levels below the Fermi energy
of this even-even nucleus lead to
the state with $I$=0, while the 2p-2h and 3p-3h particle-hole
excitation result to high-spin
toroidal isomeric states with $I_z$=26 and 40$\hbar$, respectively.

\begin{table}[h]
\begin{center}
\caption{\label{table1} The quantized values of aligned angular
  momentum $I$=$I_{z}$ for different particle-hole excitations of the
  states with $n_{z}$=0, where parameter $\Lambda_{max}$=0,1,$\ldots$
  .}
\vspace*{0.3cm}
\begin{tabular}{cll}
  \hline
  Excitation & $N$or$Z$=$4\Lambda_{max}$ & $N$or$Z$=$4\Lambda_{max}+2$ \\
  \hline
  1p-1h & $I=2\Lambda_{max}  $ & $I=2\Lambda_{max}+2 $ \\
  2p-2h & $I=4\Lambda_{max}+1$ & $I=4\Lambda_{max}+2 $ \\
  3p-3h & $I=6\Lambda_{max}  $ & $I=6\Lambda_{max}+4 $ \\
  4p-4h & $I=8\Lambda_{max}+1$ & $I=8\Lambda_{max}+4 $ \\
  5p-5h & $I=10\Lambda_{max} $ & $I=10\Lambda_{max}+6$ \\
  \hline
\end{tabular}
\end{center}
\end{table}

Table~\ref{table1} gives the simple rules to calculate
the aligned angular momentum $I$=$I_{z}$ for different p-h excitations
from states with $n_{z}$=0, for neutron number $N$ (or proton number
$Z$), such that ($N$ or $Z$ mod 4)= 0 or 2.  For nucleus $^{52}$Fe
with $N$=$Z$=26 and a parameter $\Lambda_{max}$=6, one can find from
Table~\ref{table1} the aligned angular momentum
$I_{z}$=2$\times$14, 2$\times$26, 2$\times$40, 2$\times$52,
2$\times$66$\hbar$ for 1-, 2-, 3-, 4-, 5-p-h excitations, (see also
Fig.~1(b) in Ref.~\cite{Sta14}).  With our method to study the
occurrence of toroidal high-spin isomers described above, all the
above p-h states turn out to be toroidal isomeric states as shown in
Fig.~\ref{Fig2}, with the exception of the (1p-1h)
$I_{z}$=28$\hbar$ state that is apparently below the threshold for
a non-collectively rotating toroidal nucleus.

\begin{figure}[htb]
\begin{center}
  \includegraphics[width=0.75\columnwidth]{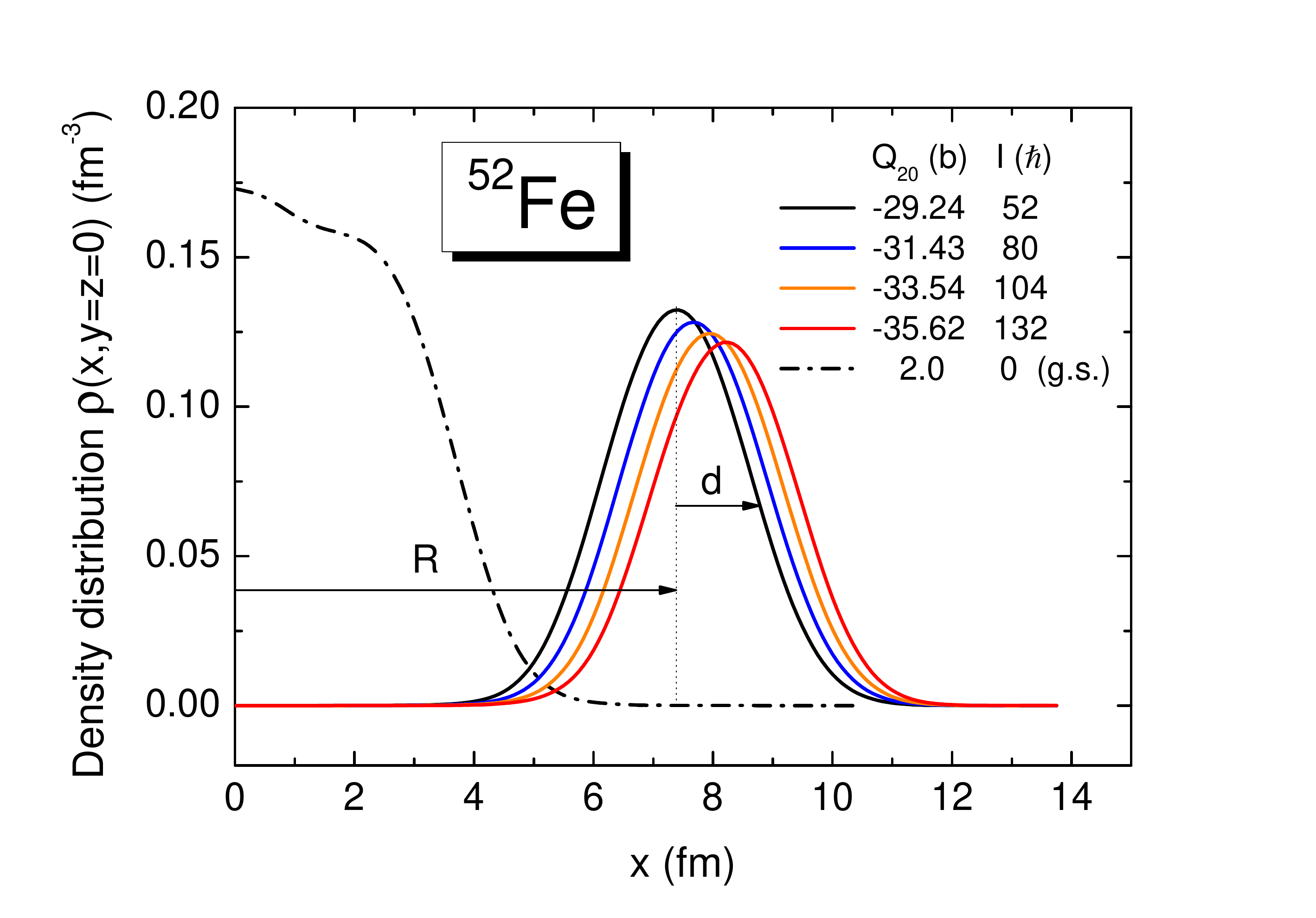}
  \caption{\label{Fig4} (Color online.) The density
    distributions of the isomeric toroidal states of $^{52}$Fe with
    $I$= 52, 80, 104, and 132$\hbar$ as a cut in the radial direction
    $x$. The values of the isomeric quadrupole moments $Q_{20}$ are
    indicated. For the first isomeric toroidal state with $I$=
    52$\hbar$ we show the major radius $R$ and the minor radius $d$
    (see text). The dash-dot curve shows the density
    distribution in the ground state of $^{52}$Fe.}
\end{center}
\end{figure}

We plot in Fig.~\ref{Fig4} the density distributions of the isomeric
toroidal states of $^{52}$Fe (presented in Fig.~\ref{Fig2}) with $I$=
52, 80, 104, and 132$\hbar$ as a cut in the radial direction $x$. One
notes that when the aligned angular momentum $I$ increases, the
maximum toroidal density $\rho_{max}$ decreases from 0.134 to 0.123
fm$^{-3}$ and the major radius $R$ increases from 7.39 to 8.20
fm. Only the minor radius $d$, defined as a half width at half maximum
(HWHM) of the toroidal distribution, stays constant at $d\approx$1.37
fm.  For comparison, we also show the total density distribution of
$^{52}$Fe in its ground state (dash-dot curve) with $\rho_{max}$=0.173
fm$^{-3}$ which is distinctly larger than $\rho_{max}$ of toroidal
distributions.  The above-mentioned features are typical for all
high-spin toroidal isomeric states of even-even, $Z$=$N$ nuclei with
28$\leq$A$\leq$52, (see Table 1 in Ref.~\cite{Sta14}).

\begin{figure}[htb]
\begin{center}
  \includegraphics[width=0.8\columnwidth]{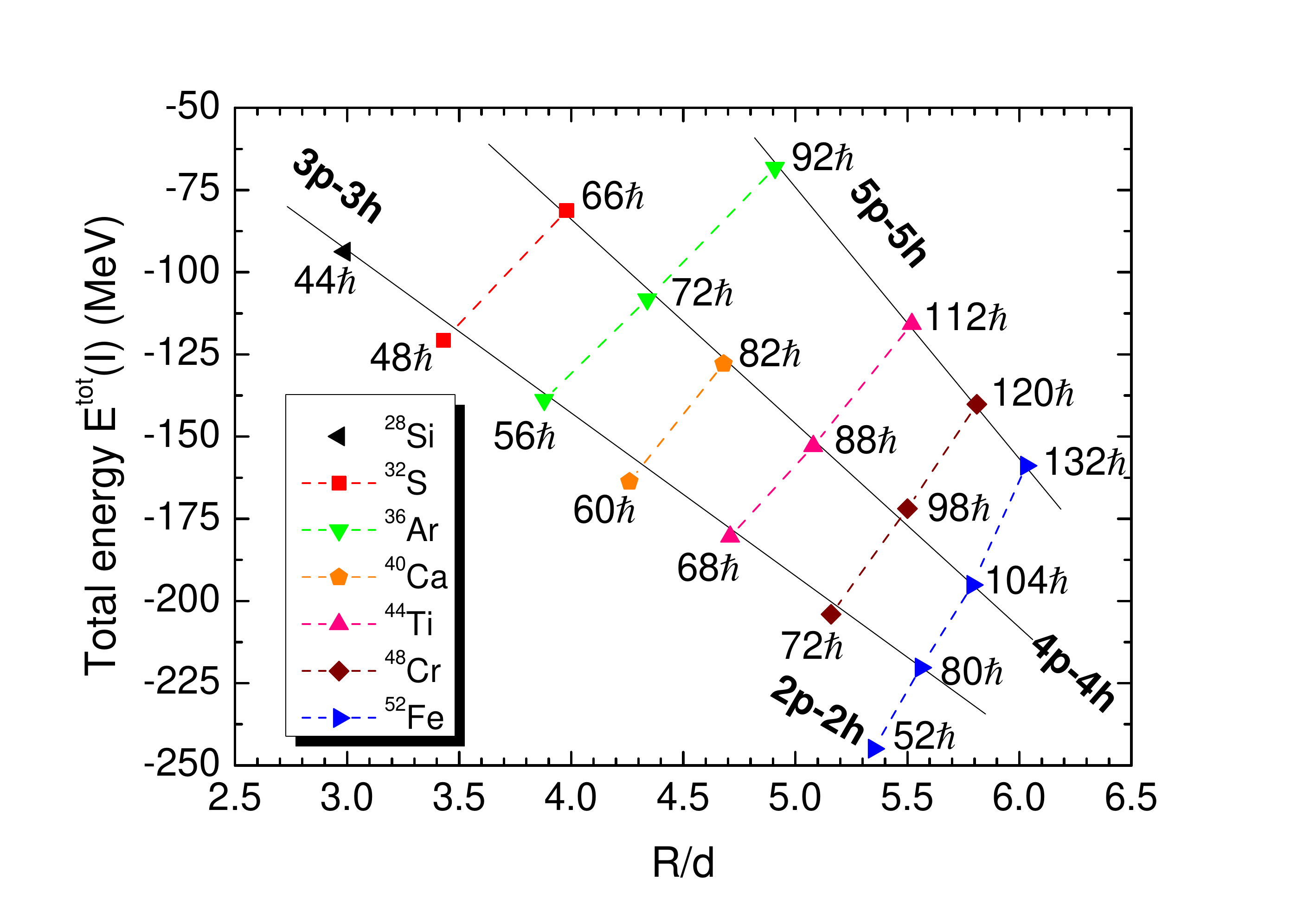}
  \caption{\label{Fig5} (Color online.) The total energy of the
    isomeric toroidal states of $^{28}$Si, $^{32}$S, $^{36}$Ar,
    $^{40}$Ca, $^{44}$Ti, $^{48}$Cr, and $^{52}$Fe of different $I$
    values as a function of $R/d$.}
\end{center}
\vspace*{-0.3cm}
\end{figure}

In Fig.~\ref{Fig5} we plot the total energy of all found isomers as a
function of $R/d$. One can see that the total energies of isomers
obtained by the same p-h excitation show linear dependence on $R/d$.
There appears to be a regular pattern on the systematics of high-spin
toroidal isomers from which properties on the nuclear fluid in the
exotic toroidal shape may be extracted.

\section{Conclusion}
Non-collective nuclear rotations with a large angular momentum
provide a favorable environment for the nuclear matter to redistribute
itself. Our exploration of the density distribution of these nuclei
with cranked self-consistent Skyrme-Hartree-Fock approach in the
region of 28$\leq$$A$$\leq$52 reveals the possibility of toroidal-shape
high-spin isomers at their local energy minima, when the angular
momenta are greater than some large thresholds. The particle-hole
nature of these non-collective rotational states, the locations of
these local energy minima, the magnitudes of the non-collective
angular momenta, and the geometrical properties of these isomeric
states have been systematically evaluated for further theoretical and
experimental explorations.

This work was supported in part by the Division of Nuclear
Physics, U.S. Department of Energy, Contract No. DE-AC05-00OR22725.

\end{document}